\documentclass[twocolumn,showkeys,preprintnumbers,amsmath,amssymb]{revtex4}

\usepackage{graphicx}
\usepackage{dcolumn}
\usepackage{bm}

\begin{document}


\title{The Carmeli metric correctly describes \\ spiral galaxy rotation curves}

\author{John G. Hartnett}
 \email{john@physics.uwa.edu.au}
\affiliation{School of Physics, the University of Western Australia\\35 Stirling Hwy, Crawley 6009 WA Australia}%


\date{\today}

\begin{abstract}
The metric by Carmeli accurately produces the Tully-Fisher type relation in spiral galaxies, a relation showing the fourth power of the rotation speed proportional to the mass of the galaxy. And therefore it is claimed that it is also no longer necessary to invoke dark matter to explain the anomalous dynamics in the arms of spiral galaxies. An analysis is presented here showing Carmeli's 5 dimensional \textit{space-time-velocity} metric can also indeed describe the rotation curves of spiral galaxies based on the properties of the metric alone.
\end{abstract}

\maketitle

\section{\label{sec:Intro}Introduction}
The rotation curves highlighted by the circular motion of stars or more accurately characterised by the spectroscopic detection of the motion of neutral hydrogen and other gases in the disk regions of spiral galaxies have caused concern for astronomers for many decades. Newton's law of gravitation predicts much lower orbital speeds than those measured in the disk regions of spiral galaxies. Most luminous galaxies show slightly declining rotation curves (orbital speed vs radial position from nucleus) in the regions outside the star bearing disk, coming down from a broad maximum in the disk. Intermediate mass galaxies have mostly nearly flat rotation speeds along the disk radius. Lower luminosity galaxies usually have monotonically increasing orbital velocities across the disk. (See \cite{Sofue2001} for a good review.) The traditional solution has been to invoke halo `dark matter'   \cite{Begeman1991} that surrounds the galaxy but is transparent to all forms of electromagnetic radiation. In fact, astronomers have traditionally resorted to `dark matter' whenever known laws of physics were unable to explain the observed dynamics.

In 1983 Milgrom introduced his MOND \cite{Milgrom1983a,Milgrom1983b,Milgrom1983c}, an empirical approach, which attempts to modify Newtonian dynamics in the region of very low acceleration. Newton's law describes a force proportional to $r^{-2}$, where $r$ is the radial position, but Milgrom finds that a $r^{-1}$ law fits the data very well \cite{Begeman1991}. Others have also attempted to formulate modified force laws, such as Disney \cite{Disney1984}, Wright \cite{Wright1990} and Carmeli \cite{Carmeli2000,Carmeli2002}. The latter formulated a modification and an extension of Einstein's general theory, in an expanding universe taking into account the Hubble expansion, which imposes an additional constraint on the dynamics of particles \cite{Carmeli1982}. 

Carmeli believes the usual assumptions in deriving Newton's gravitational force law from general relativity are insufficient, that gases and stars in the arms of spiral galaxies are not immune from Hubble flow. As a consequence a universal constant $a_{0}$ (in this case, slightly different to Milgrom's) is introduced as the minimum acceleration in the cosmos.  

Using this theory Carmeli \cite{Carmeli1998} successfully provided a theoretical description of the Tully-Fisher law, deriving the proportionality between the fourth power of the galaxy's rotation speed ($v$) and its mass ($M$), vis-$\grave{a}$-vis 

\begin{equation} \label{eqn:TF}
\ v^{4}=\frac{2} {3} G M a_{0},
\end{equation}
\\
where $G$ is the Newtonian gravitational constant. 

Equation (\ref{eqn:TF}) can be re-written as $v^{4} =  \left(\frac {G M} {r^{2}} \frac{2} {3} a_{0}\right)  r^{2}$ and therefore the positive square root is $v^{2}/r = \sqrt{ g_{N} \frac{2} {3} a_{0}}$, where $g_{N}$ is the Newtonian gravitational acceleration. Hence the latter is consistent with Milgrom's phenomenological approach in the low acceleration limit.

\section{\label{sec:5Dcosmology}Carmeli's 5D cosmology}

In the weak gravitational limit, where Newtonian gravitation applies, it is sufficient to assume the Carmeli metric with non-zero elements $g_{00} = 1 + \phi /c^{2}$, $g_{44} = 1 + \psi /\tau^{2}$, $g_{kk} = -1$, ($k = 1, 2, 3$) in the lowest approximations in both $1/c$ and $1/\tau $. Here a new constant, called the Hubble-Carmeli constant, is introduced $\tau \approx 1/H_{0}$ . The potential functions $\phi$ and $\psi$ are determined by Einstein's field equations and from their respective Poisson equation, either

\begin{equation} \label{eqn:Poisson}
\nabla^{2} \phi = 4 \pi G \rho_{m} \;  or \;  \nabla^{2} \psi = \frac {4 \pi G} {a_{0}^{2}} \rho_{m},
\end{equation}
\\
where  $\rho_{m}$ is the mass density and $a_{0}$ the universal characteristic acceleration $a_{0} =c/\tau$. As usual $c$ is the speed of light in vacuo.

The Hubble law describes the expansion of the cosmos and the matter embedded in it. Therefore the line element for any two points in this `new' \textit{space-time-velocity} is $ds^{2} = g_{00} c^{2} dt^{2}+ g_{kk} dx^{k}+ g_{44} \tau^{2} dv^{2} = 0$. The relative separation in 3 spatial coordinates $r^{2} = (x^{1})^{2} + (x^{2})^{2} + (x^{3})^{2}$ and the relative velocity between points connected by $ds$ is $v$. The Hubble-Carmeli constant,  $\tau$, is a constant for all observers at the same epoch, therefore may be regarded as a constant on the scale of any measurements.

The equations of motion (B.62a and B.63a from \cite{Carmeli2002}) to lowest approximation in $1/c$ are reproduced here,

\begin{equation} \label{eqn:derivphi}
\frac{\partial^{2} x^{k}} {\partial t^{2}} = - \frac {1} {2} \frac{\partial \phi} {\partial x^{k}}.
\end{equation}
\\
This is the usual looking geodesic equation derived from general relativity but now in 5 dimensions. And the second is a new phase space equation derived from the theory,

\begin{equation} \label{eqn:derivpsi}
\frac{\partial^{2} x^{k}} {\partial v^{2}} = - \frac {1} {2} \frac{\partial \psi} {\partial x^{k}}.
\end{equation}
\\
The solution of Einstein's field equations in 5D result in 

\begin{subequations}
\begin{eqnarray}
\phi = \frac{1-\Omega} {\tau^{2}}r^{2} - \frac {2 G M} {r},  \label{eqn:phi}
\\
\psi = \frac{1-\Omega} {c^{2}}r^{2} - \frac {2 G M} {r} \frac {\tau^{2}} {c^{2}}, \label{eqn:psi}
\end{eqnarray}
\end{subequations}
\\
where  $\Omega$ is the matter density expressed as a fraction of the critical or `closure' density, which in this model is defined by $\rho_{c} = 3/8 \pi G \tau^{2}$.

From (\ref{eqn:derivphi}) and (\ref{eqn:phi}) it follows  

\begin{equation} \label{eqn:accel}
\frac {\partial^{2} x^{k}} {\partial t^{2}} = - \frac{1-\Omega} {2 \tau^{2}}(r^{2}),_{k} + G M \left(\frac {1} {r}\right),_{k} .
\end{equation}
\\

\subsection{\label{sec:postNewtonian}Post-Newtonian force law}
By carrying out the differentiation with respect to $r$, (\ref{eqn:accel}) becomes a new post-Newtonian force equation

\begin{equation} \label{eqn:postN}
g(r) = - \frac{1-\Omega} { \tau^{2}}r -  \frac {G M} {r^{2}}.
\end{equation}
\\

For $\Omega$  less than critical ($\Omega < 1$) in or near a galaxy this means an additional force inwards is applied to the test particle. For $\Omega$  more than critical ($\Omega  > 1$) it represents an additional outward force. The solution of (\ref{eqn:postN}) for small $r$ is the familiar Newtonian equation. The first term on the right-hand side of (\ref{eqn:postN}) Carmeli neglects as small on a galaxy scale.  But that would only be true if the matter density ($\Omega$) is a descriptor of space curvature on a much larger scale than the galaxy.  If it describes the local density then the term is not insignificant.

\subsection{\label{sec:circular}Post-Newtonian circular motion}
From (\ref{eqn:derivpsi}) and (\ref{eqn:psi}) it follows that

\begin{equation} \label{eqn:phasespace}
\frac {\partial^{2} x^{k}} {\partial v^{2}} = - \frac{1-\Omega} {2 c^{2}}(r^{2}),_{k} + \frac{G M \tau^{2}} {c^{2}} \left(\frac {1} {r}\right),_{k} .
\end{equation}
\\

Integrating with respect to $dx^{k}$ (\ref{eqn:phasespace}) becomes

\begin{equation} \label{eqn:phasespace2}
\left (\frac {dr} {dv} \right)^{2} = \frac{\Omega-1} {2 c^{2}}r^{2} + \frac{G M \tau^{2}} {c^{2}} \frac {1} {r},
\end{equation}
\\
which is the new post-Newtonian equation for circular motion.  

For $r$ small, (\ref{eqn:phasespace2}) becomes equation (8) of \cite{Carmeli1998}. Because the new dimension ($v$) in the Carmeli metric is constructed as an analogue to the time co-ordinate in the usual 4D \textit{spacetime}, the meaning of the new equation is identified by the substitutions $v \rightarrow t$, $\tau \rightarrow c$. In this case, with these substitutions, and  $\Omega = 1$ representing Euclidean space, we recover the usual equation for circular motion. 

\subsection{\label{sec:Ontology}Ontology}
A possible ontology for (\ref{eqn:phasespace2}) is that even though galaxies are constrained by gravity against expansion in the radial direction, they are free to expand tangentially to the radius i.e. azimuthally in the plane of the galaxy. 

It is known that some galaxies have portions of their disks, and the extended gas regions beyond, that rotate in opposite directions to each other \cite{Sofue2001}. From the suggested ontology, it would be expected that this situation could occur. As a new galaxy develops the direction of rotation is determined by the initial angular momentum of the system. For proto-galaxies with low initial angular momentum different portions of the developing disk can rotate in response to the Hubble flow force in the azimuthal direction, some parts prograde, some parts retrograde.

\subsection{\label{sec:Phasespaceequation}Phase space equation}
After integrating (\ref{eqn:phasespace2}) and solving for radial distance ($r$) as a function of velocity ($v$) with the condition    $v(r = 0) = 0$, we get 

\begin{equation} \label{eqn:rvsv}
r = \left(\frac{2 G M \tau^{2}} {\Omega - 1} \right)^{\frac {1} {3}} sinh^{\frac {2} {3}}\left( \frac {3} {2} \frac {v} {c}  \sqrt{ \frac {\Omega - 1} {2}}\right),
\end{equation}
\\
which, to lowest order in $v/c$, reduces to  

\begin{equation} \label{eqn:rvsvapprox}
r = \left(\frac{3} {2} \right)^{\frac {2} {3}} \left(\frac {G M \tau^{2}} {c^{2}}\right)^ \frac {1} {3} v^{\frac{2} {3}}.
\end{equation}
\\
This is equation (9) of \cite{Carmeli1998}. For typical-mass galaxies (\ref{eqn:rvsvapprox}) describes orbital speeds of hundreds of kilometres per second. On small $r$ scales, or scales of small local orbital motion, the effect of (\ref{eqn:rvsvapprox}) is small. Assuming $r$ small so the first term is negligible and integrating (\ref{eqn:postN}) with respect to $dr$ we get the familiar equation for circular motion 

\begin{equation} \label{eqn:normalcircmotion}
v^{2} = \frac {G M} {r},
\end{equation}
\\
where $v$ is the circular or tangential velocity in the system. Simultaneously solving (\ref{eqn:normalcircmotion}) and (\ref{eqn:rvsvapprox}) by eliminating $r$ yields a Tully-Fisher type relation as in (\ref{eqn:TF}). This is the result of \cite{Carmeli1998}.

\section{\label{sec:Galaxyrotationcurves}Galaxy rotation curves}
The first term in (\ref{eqn:phi}) and hence (\ref{eqn:postN}) is valid where $ \phi/c^{2} \ll 1$. By substituting $\rho_{c}$ into (\ref{eqn:phi}) and assuming $\Omega \gg 1$ it can be shown this condition is satisfied when $r^{2} \rho_{m} < 10^{27}$ (in SI units), where $\rho_{m}$ is the local baryonic matter density. For radial distances within a galaxy ($r$) on the scale of a few $kpc$s ($1 \, kpc \approx 3 \times 10^{19}\, m$) the density of material  $\rho_{m} < 10^{-14} \; g \; cm^{-3}$. This is true for most regions in a galaxy; the exception would be near a compact object. 

By rearranging (\ref{eqn:postN}) using $\Omega = \rho_{m}/\rho_{c}$ where $\Omega \gg 1$ we get

\begin{equation} \label{eqn:postNg}
g(r) = \frac{8 \pi G \rho_{m}} {3}r -  \frac {G M} {r^{2}}.
\end{equation}
\\
Note there is no term dependent on the Hubble-Carmeli time constant ($\tau$). Equation (\ref{eqn:postNg}) is a total local force law, and a post-Newtonian equation. It needs significant investigation on the scale of galaxies, clusters etc. It indicates a gravity shielding type effect by the matter density field. For small radii from a central gravitating body, the first term is insignificant with respect to the second term and (\ref{eqn:postNg}) becomes the familiar Newtonian equation. For the Sun-Earth system the second term is about 30 orders of magnitude larger than the first.

For   $\Omega \gg 1$, (\ref{eqn:phasespace2}) becomes

\begin{equation} \label{eqn:phasespace3}
\left (\frac {dr} {dv} \right)^{2} = \frac{4 \pi G \rho_{m}} {3 a_{0}^{2}}r^{2} + \frac{G M} {a_{0}^{2}} \frac {1} {r}.
\end{equation}
\\
Notice $a_{0}$ (= $c/\tau$) explicitly appears in this equation. In the limit of zero mass and zero distance (\ref{eqn:phasespace2}) can be re-written as
 
\begin{equation} \label{eqn:zeromassHL}
\left (\frac {dr} {dv} \right)^{2} = \tau^{2},
\end{equation}
\\
which is the Hubble relation. Each term on the rhs of (\ref{eqn:phasespace3}) contributes half the gravitational radius cancelling the radius in the denominator.

Now (\ref{eqn:phasespace3}), like (\ref{eqn:postNg}), is galaxy specific and depends on the matter density in the disk. Using (\ref{eqn:postNg}) a new expression for the speed ($v$) of circular motion of stars and gases will be derived as a function of distance ($r$) from the centre of a galaxy. Real values of typical densities will be used to see if a connection can be found to the Milgrom formula, which means $g(r) \rightarrow \sqrt{g_{N} a_{0}}$ as $r \rightarrow \infty$. In this case it turns out $a_{0}$ is different by a constant factor. 

By substitution of $\Omega - 1 \rightarrow 8 \pi G \tau^{2} \rho_{m}/3 $  into (\ref{eqn:rvsv}), for  $\Omega \gg 1$, we get

\begin{equation} \label{eqn:rapprox}
r = \left(\frac{3 M}{4 \pi \rho_{m}} \right)^{\frac {1} {3}} sinh^{\frac{2} {3}}\left(\frac{v} {c} \sqrt{3 \pi G \rho_{m} \tau^{2}}  \right).
\end{equation}
\\
By integrating (\ref{eqn:postNg}) with respect to $dr$ we obtain the new equation for circular motion

\begin{equation} \label{eqn:newcircmotion}
v^{2} = \frac{4 \pi G \rho_{m}}{3} r^{2} + \frac{G M} {r} 
\end{equation}
for $\Omega \gg 1$. Notice this equation contains a new term, which adds additional velocity to the stars and gases circulating in the disk region of the galaxy. But for small $r$ the term is insignificant. 

If we initially assume that the argument of the hyperbolic sine function in (\ref{eqn:rapprox}) is much less than unity, that is,

\begin{equation} \label{eqn:approx1}
\frac {v} {c} \sqrt{3 \pi G \rho_{m} \tau^{2}} = x \ll 1
\end{equation}
which implies $v \ll 150 \; km \; s^{-1}$ for  $\rho_{m} = 10^{-23} g \; cm^{-3}$. This is reasonable in some disk regions of galaxies, but when higher velocities are measured this approximation cannot be used. 

Using the expansion $sinh^{\frac {2} {3}} (x) \approx x^{\frac {2} {3}} + x^{\frac{8}{3}}/9$ for $x \ll 1$, collecting terms lowest in $x$ and substituting $r$ from (\ref{eqn:rapprox}) into (\ref{eqn:newcircmotion}) for $v \ll c$ for circular motion, we get

\begin{equation} \label{eqn:vsqrdapprox}
v^{2} = \sqrt{\frac{2}{3} G M a_{0}}\left(1 + \frac{3 \pi G \rho_{m}}{a_{0}^{2}} v^{2} \right)^{\frac{3}{4}}.
\end{equation}

For very low matter density ($\rho_{m}\rightarrow 0$) this becomes the Tully-Fisher type law for galaxies as in (\ref{eqn:TF}). However where the approximation (\ref{eqn:approx1}) is true, it also follows that
\begin{equation} \label{eqn:approx2}
\frac{3 \pi G \rho_{m}}{a_{0}^{2}} v^{2} = \left(\frac{v}{c} \sqrt{3 \pi G \rho_{m} \tau^{2}}\right)^{2} \ll 1.
\end{equation}
\\
Hence (\ref{eqn:vsqrdapprox}) becomes

\begin{equation} \label{eqn:vsqrdapprox2}
v^{2} = \sqrt{\frac{2}{3} G M a_{0}}\left(1 + \frac{3}{4} \frac{3 \pi G \rho_{m}}{a_{0}^{2}} v^{2} \right)
\end{equation}
and the solution for circular velocity is simply found.

\begin{equation} \label{eqn:vsqrdapprox2}
v^{2} = \sqrt{\frac{2}{3} G M a_{0}}\left(1 - \frac{9 \pi G \rho_{m}}{4 a_{0}^{2}} \sqrt{\frac{2}{3} G M a_{0}}. \right)^{-1}
\end{equation}

Equation (\ref{eqn:vsqrdapprox2}) applies in all regions of spiral galaxies with densities that $10^{-30} \; g \; cm^{-3} < \rho_{m} < 10^{-23} \; g \; cm^{-3}$. Densities in this range yield circular velocities in the disk regions like $85 \; km \; s^{-1}$ for galaxies with mass $M = 10^{9} M_{\odot}$ and $150 \; km \;s^{-1}$ for galaxies with mass $M = 10^{10} M_{\odot}$ (where $M_{\odot}$ is a solar mass unit = $2 \times 10^{30}\; kg$). For larger masses, the higher velocities mean the approximation is invalid. Therefore for higher velocities and/or densities  $\rho_{m} > 10^{-23}\; g \; cm^{-3}$ (\ref{eqn:rapprox}) must be used without approximation. 

\section{\label{sec:MOND}Milgrom's MOND}
We can write  $g_{M}(r) = v^{2}/r $ where $g_{M}(r)$ is the Milgrom acceleration inferred from circular motion and therefore it follows from (\ref{eqn:vsqrdapprox2}) that

\begin{equation} \label{eqn:MONDapprox}
g_{M}(r) = \sqrt{\frac{2}{3} g_{N}(r) a_{0}}\left(1 - \frac{9 \pi G \rho_{m}}{4 a_{0}^{2}} \sqrt{\frac{2}{3} G M(r) a_{0}} \right)^{-1},
\end{equation}
where $M(r)$ is the enclosed mass at radius $r$. By squaring (\ref{eqn:MONDapprox}) it follows
\begin{equation} \label{eqn:MONDsqr}
g_{M}^{\;2}(r) = g_{N}(r) a_{0}'\left(1 - \frac{\pi G \rho_{m}}{(a_{0}')^{2}} \sqrt{\frac{2}{3} G M(r) a_{0}'} \right)^{-2},
\end{equation}
where $g_{N}(r)$ is the Newtonian acceleration and $a_{0}' = \frac{2}{3} a_{0} \approx 5.2 \times  10^{-10} \; m\; s^{-2}$, depending on the  precise value of $\tau$.

In the region of low acceleration on the edge of a galaxy where $\rho_{m} \rightarrow 0$ as $r \rightarrow$   large, and $M(r) \rightarrow$ a constant, (\ref{eqn:MONDsqr}) becomes
\begin{equation} \label{eqn:MONDsqrlimit}
g_{M}^{\;2}(r) = g_{N}(r) a_{0}'
\end{equation}
which is the low acceleration limit of Milgrom's phenomenological law.

\section{\label{sec:Toymodel}Toy model: density profile}
Let us assume a simple density distribution for a spiral galaxy valid in the disk region, expressed in cylindrical co-ordinates $(r, \phi , z)$ but with no $\phi$-dependence on mass density 
\begin{equation} \label{eqn:rho}
\rho_{m}(r,z) = A r^{-\alpha} e^{-\beta z},
\end{equation}
\\
where $A, \alpha$ and $\beta$ are parameters to be determined. Here $\beta  = 1/z_{0}$, and $z_{0}$ is an exponential scale factor of order of several $pc$ for a galaxy. Therefore in this simple toy model the accumulated mass of the galaxy $M(r)$ can be written as the sum of two parts, the bulge mass ($M_{0}$) plus the disk mass, hence
\begin{eqnarray} \label{eqn:galmass}
M(r)&= M_{0} + 2 \pi \int^{r}_{0} rdr \int^{z}_{-z} dz \; \rho_{m}(r,z) \nonumber \\
&= M_{0} + 2 \pi A \int^{r}_{0} r^{1-\alpha} dr \; 2\int^{z}_{0}e^{-z/z_{0}} dz,
\end{eqnarray}
\\
which yields mass as a function of $r$.
\begin{equation} \label{eqn:galmass2}
M(r) = M_{0} + \frac {4 \pi A z_{0}} {2-\alpha} r^{2-\alpha},
\end{equation}
\\
where $ \alpha  \neq 2$.

Also assume the density is uniform in $z$ and valid for $z = 0$, hence in the appropriate units,
\begin{equation} \label{eqn:rho2}
\rho_{m}(r) = A r^{-\alpha}
\end{equation}
\\

Substituting (\ref{eqn:galmass}) and (\ref{eqn:rho2}) into (\ref{eqn:vsqrdapprox2}) we get for circular motion
\begin{eqnarray} \label{eqn:circ3}
v^{2} = \sqrt{\frac{2}{3} G \left(M_{0} + \frac {4 \pi A z_{0}} {2-\alpha} r^{2-\alpha}\right)a_{0}}\times \nonumber \\
\left(1-\frac{9 \pi G A r^{-\alpha}}{4 a^{2}_{0}}\sqrt{\frac{2}{3} G \left(M_{0} + \frac {4 \pi A z_{0}} {2-\alpha} r^{2-\alpha}\right)a_{0}}\right)^{-1}.
\end{eqnarray}

\begin{figure}
\includegraphics[width = 3.5 in]{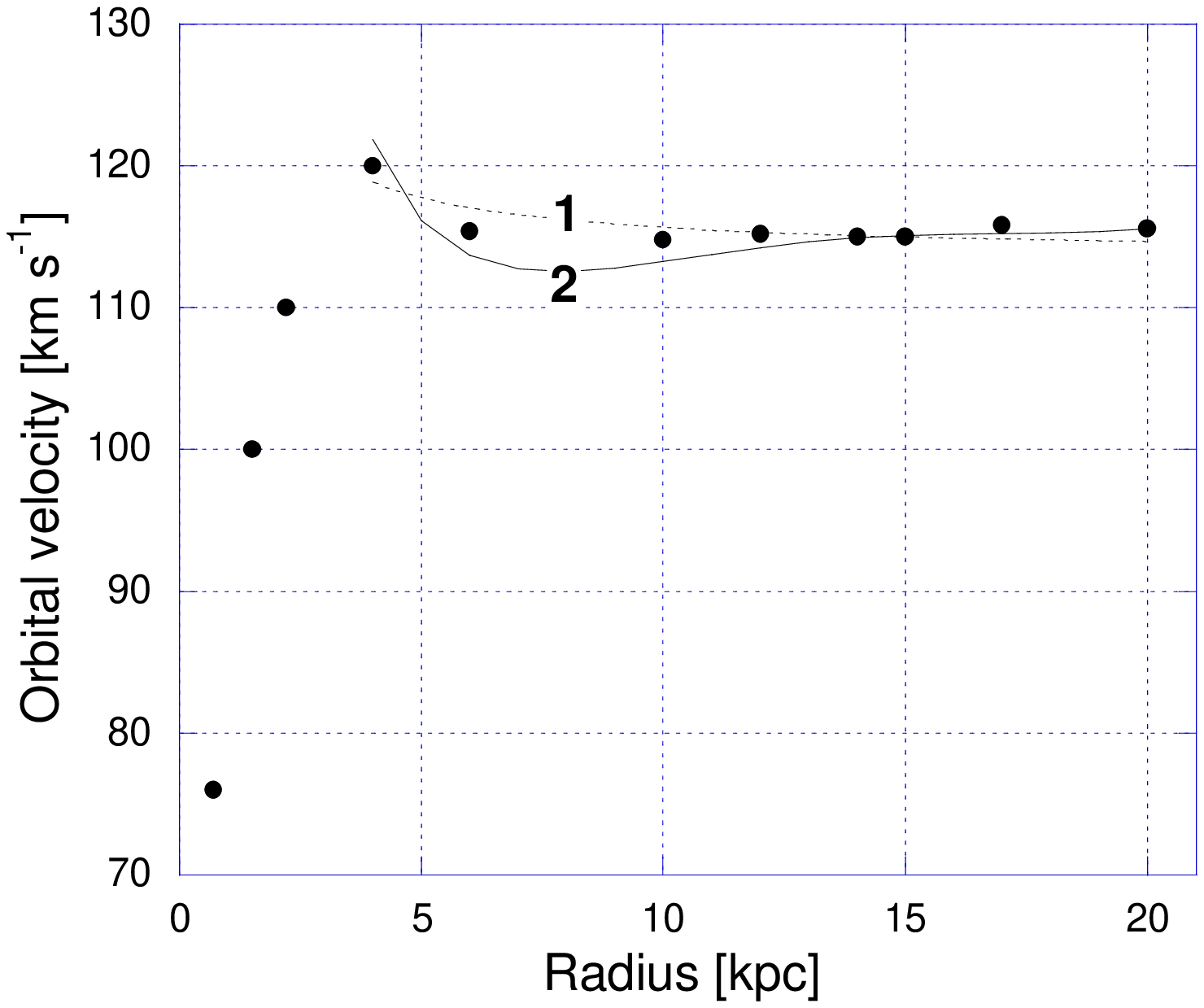}
\caption{\label{fig:fig1} Fit to the velocity-distance data of the spiral galaxy NGC6503. Filled circles are the data, and the dashed line (curve 1) the fit from (\ref{eqn:general}). The parameters $M_{0} = 1.836 \times 10^{9} M_{\odot}$ and $\alpha  = 1.0075$ were determined from the fit. Curve 2 (solid line) represents the fit using an undulating density function as shown in figure \ref{fig:fig3}}
\end{figure}

Assuming $z_{0} = 1 \, pc$, (\ref{eqn:circ3}) was fit to the velocity-distance data of NGC6503 from \cite{Begeman1991} with the results shown in  curve 1 (dashed line) in figure \ref{fig:fig1}. The best fit parameters for $M_{0}$, $A$ and $\alpha$  were determined. The fit yielded a first order density dependence on $r$. It also showed a good agreement, showing the correct trend, especially in the disk region, considering the simple model used.  The total mass of the galaxy is then $M(r = 15 \; kpc)= M_{0} + 4\pi A z_{0} r  = (1.836 + 0.165) \times 10^{9} \; M_{\odot}$. Therefore the disk comprises $8.9 \%$ of the total galaxy mass and the total mass of the galaxy equals $2.0 \times 10^{9}\; M_{\odot}$.  This compares with published value of $4.8 \times 10^{9} \; M_{\odot}$ \cite{Begeman1991} for the luminosity mass of the galaxy. It is underestimated, which is the reverse of the dark matter problem. 

But most importantly no dark matter is needed to explain the rotation curve. The shape of the derived dependence of the tangential speed of the stars to their radial distance from the centre of the galaxy is in accordance with expectation. A more comprehensive model though is needed especially for the region close to the central bulge. That should involve a gas component as well as a more precise model of the enclosed mass. Also the approximate model used here would not apply as the density and orbital speeds will be greater in many cases. However the purpose of the toy model presented here is to show that it can work in the outer disk region of a galaxy and realistic though much smaller masses result from the fit than when halo dark matter is invoked.

\section{\label{sec:Generalsolution}General solution}
After substituting (\ref{eqn:rapprox}) into (\ref{eqn:newcircmotion}), eliminating $r$, a transcendental equation of the following form results, 

\begin{equation} \label{eqn:general}
\delta v^{2} sinh^{2/3}(\varsigma v) - \gamma sinh^{2}(\varsigma v) - \gamma = 0,
\end{equation}
\\
where $\delta = \left( 3 M/4 \pi \rho_{m}  \right)^{1/3}$, $\varsigma = \sqrt{3 \pi G \rho_{m}}/a_{0}$, $\gamma = G M$.

\begin{figure}
\includegraphics[width = 3.5 in]{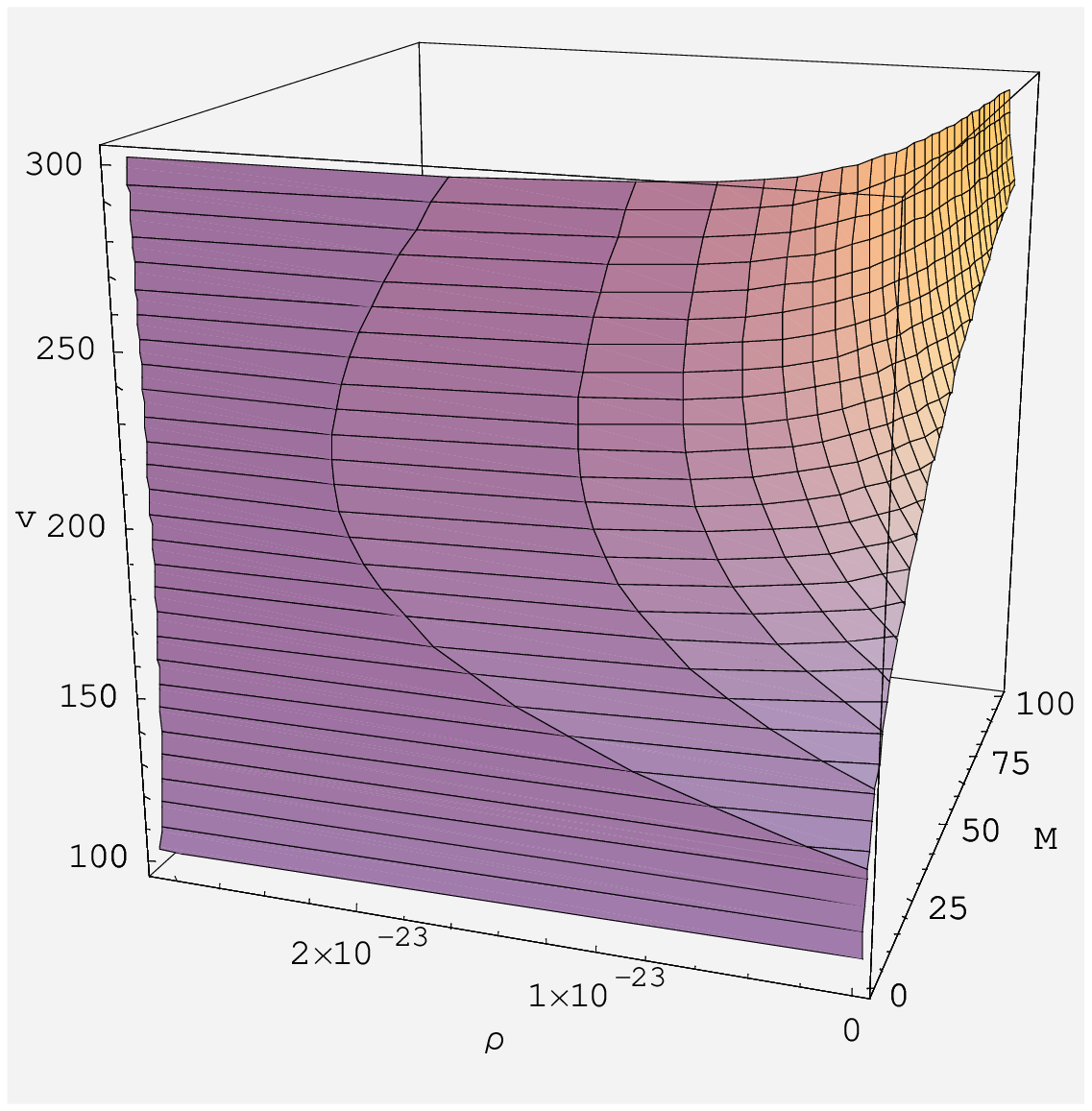}
\caption{\label{fig:fig2} Solution of (\ref{eqn:general}) with orbital velocity ($v$) on the vertical axis in $km \; s^{-1}$, the galaxy mass ($M$) in $10^{9} M_{\odot}$ units and the matter density ($\rho_{m}$) in $g \; cm^{-3}$}
\end{figure}

Equation (\ref{eqn:general}) must be solved for the motions of stars and gases in any region of a galaxy.  An exact solution of (\ref{eqn:general}) is of course not analytically possible but the solution may be visualized using a 3D plot generated by the Mathematica software package. This is shown in figure \ref{fig:fig2}. It is apparent that as we follow a line of constant mass ($M$) we see the tangential velocity ($v$) increase as a function of local matter density ($\rho_{m}$). The more massive the galaxy the less sensitive the curves are to changes in density.  

\begin{figure}
\includegraphics[width = 3.5 in]{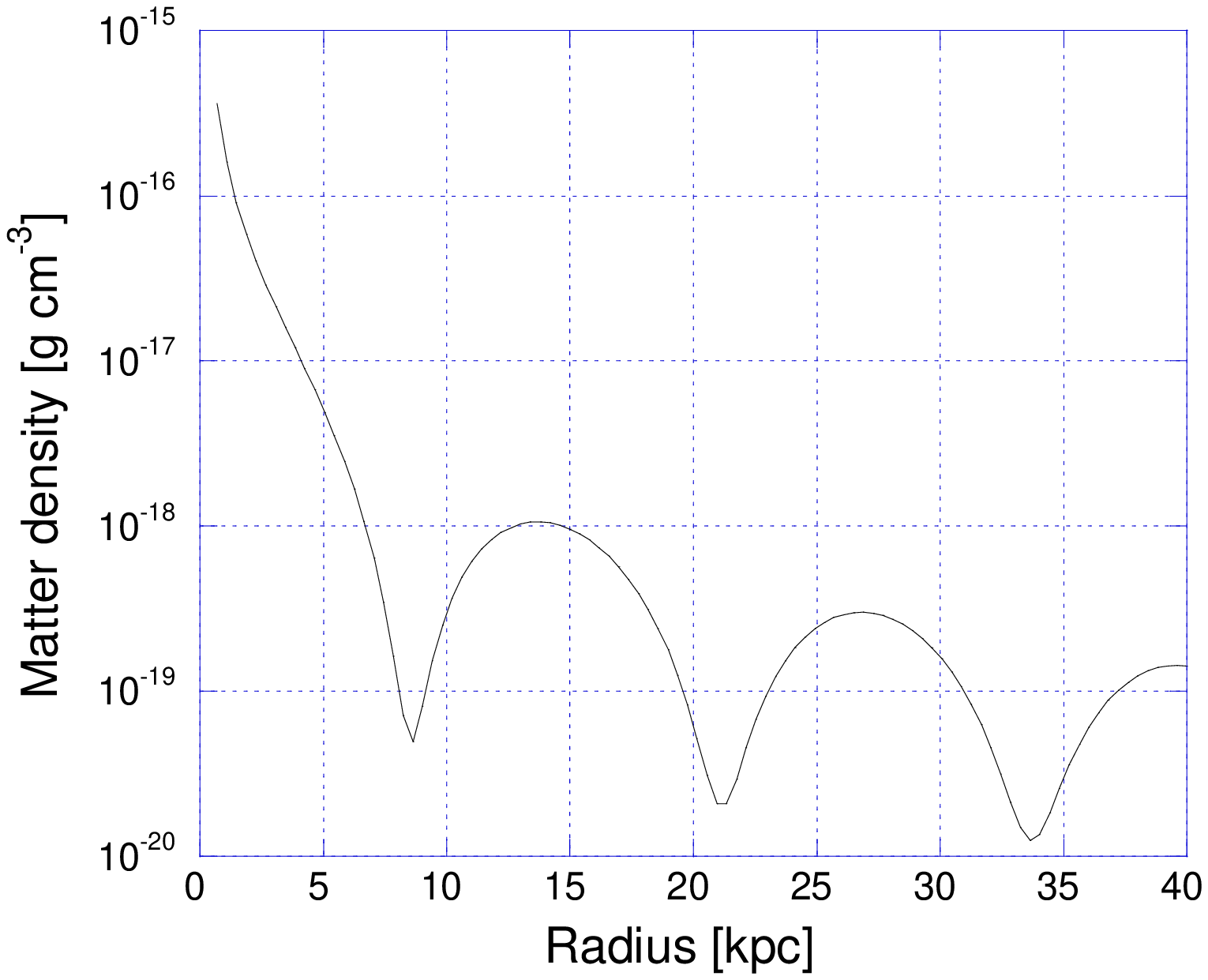}
\caption{\label{fig:fig3} Simulated density fluctuations (in $g \; cm^{-3}$) as a function of disk radius (in $kpc$)}
\end{figure}

\subsection{\label{sec:Densityfluctuations}Density fluctuations}
As is noted in many galaxy rotation curves the velocity often rises and falls as a function of radial distance. This is indicative of changes in density in the disk region. In order to simulate the undulating density fluctuations in the arms of a spiral galaxy, which commonly are of the order of a few tens of kilometres per second \cite{Sofue2001}, I have chosen a density function as shown in figure \ref{fig:fig3}. This function has a mean $r^{-1}$ dependence on radius as per the result of the toy model above. Using this function and a mass derived from (\ref{eqn:galmass2}), replacing the density function, (\ref{eqn:general}) was solved for 20 fixed points along the radius, where mass and density are both functions of radius. The mass $M_{0}$ was adjusted to get the best fit to the simulated data of NGC6503 and the result is shown in  curve 2 (solid line) in figure \ref{fig:fig1}. The mass of the galaxy was then determined to be $3.2 \times 10^{9} M_{\odot}$ with $14 \%$ of the mass in the disk out to $15 \; kpc$. This is somewhat different to the toy model result above, and the fit is not as good, however I have tried to simulate the effect of density fluctuations. Note it has not been attempted to simulate density and mass within the central $5 \; kpc$. That will be left to future work.

\begin{figure}
\includegraphics[width = 3.5 in]{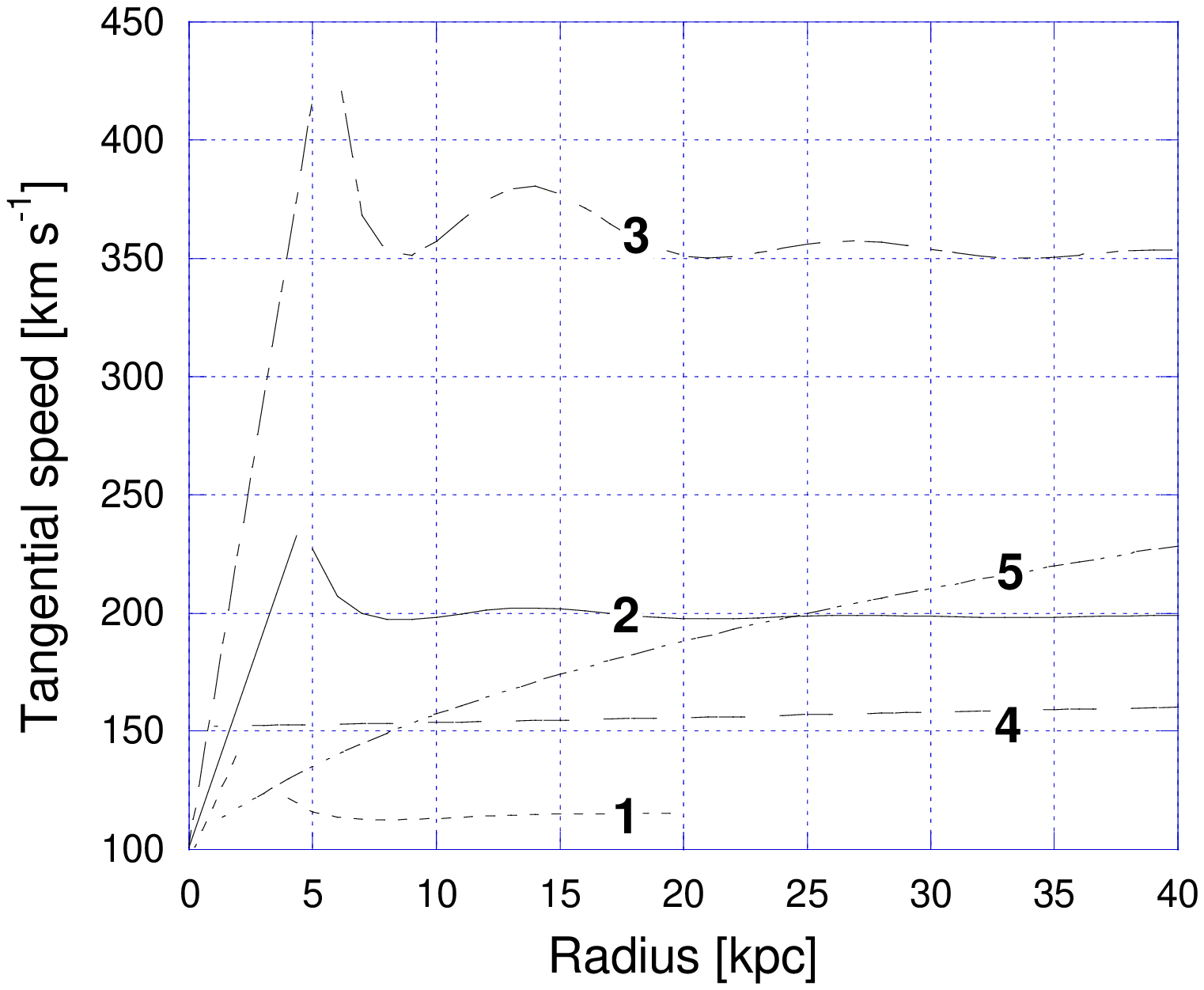}
\caption{\label{fig:fig4} Various rotation curves generated from (\ref{eqn:general}). See the text for details}
\end{figure}

By changing the mass of the central nuclear bulge from $M_{0} = 2.8 \times 10^{9} M_{\odot}$ by 1 and 2 orders of magnitude the following curves in figure \ref{fig:fig4} were generated. Curve 1 is the same curve as curve 2 in figure \ref{fig:fig1}, curve 2 is with $M_{0} = 28  \times 10^{9} M_{\odot}$ and curve 3 is with $M_{0} = 280 \times 10^{9} M_{\odot}$ in the central core. Notice the stronger density fluctuations and the rigid body steep nuclear rise. Normally the latter would fall well within a $kpc$ but I have not attempted to model the nuclear bulge region.

In each of the above the density dependence was $r^{-1}$ (i.e. $\alpha = 1$) and with about $10 \%$  of the mass in the disk out to $15 \; kpc$, as in the toy model above. By altering the density fall off we get a rising velocity curve. Curve 4 of figure \ref{fig:fig4} represents a galaxy with a $10^{10} M_{\odot}$ core and only $7.5 \%$ in the disk out to $15 \; kpc$ and a density function falling as $r^{-0.9}$  (i.e. $\alpha = 0.9$). Curve 5 represents a galaxy with a density dependence of $r^{-0.8}$ (i.e. $\alpha = 0.8$) and $M_{0} = 2.8 \times 10^{9} M_{\odot}$ but with $14.4 \times 10^{9} M_{\odot}$ in the disk out to $15 \; kpc$. That is $83\%$ in the disk. This curve shows a continuous rise as a function of distance as most of the mass is exterior to the nucleus.

\subsection{\label{sec:Galaxymass}Galaxy mass}
Equation (\ref{eqn:general}) can be rearranged to calculate the mass of a galaxy from the known velocity and an estimate on the density.

\begin{equation} \label{eqn:mass}
M = \left(\frac{3}{4 \pi G^{3} \rho_{m}} \right)^{1/2}  \frac{v^{3} sinh(\varsigma v)}{cosh^{3}(\varsigma v)}. 
\end{equation}

Equation (\ref{eqn:mass}) is a post-Newtonian equation and contains, besides constants, a measured velocity and the matter density which needs to be estimated in the region where the velocity is measured. The universal acceleration constant $a_{0}$ is contained in $\varsigma$.  Here it was assumed that $a_{0} = 7.8 \times 10^{-10} \; m \; s^{-2}$ after Carmeli \cite{Carmeli2002}. 

Using the measured orbital speeds around the Milky Way galactic centre we can calculate an enclosed mass from (\ref{eqn:mass}) and compare it to the enclosed mass derived from the Newtonian circular motion equation (\ref{eqn:normalcircmotion}) as well as the commonly cited mass of $10^{11} M_{\odot}$ for the Galaxy.  Table \ref{tab:table1} shows the results of the calculations. Our sun's position is within the $9.95 \; kpc$ and the enclosed mass is 4 times smaller from the post-Newtonian equation when a reasonable density is assumed. If a density 2 orders of magnitude larger is assumed, the mass is 7 times smaller. At $18.4 \; kpc$ the enclosed mass is 3 times smaller depending on the matter density. 

From Table \ref{tab:table1} is can be clearly seen that the `missing' mass or the dark matter that is supposed to haunt the halo regions of our galaxy is the result of the incorrect equation of circular motion. Depending on the matter density distribution through the galaxy the total mass may in fact be at least 4 times less than previously calculated.

\begin{table}
\caption{\label{tab:table1}Expected mass of Milky Way galaxy}
\begin{tabular}{ccccc}
\hline
 Mass from & Mass from & Density & Radial position & Velocity \\
 (\ref{eqn:normalcircmotion})[$10^{9} M_{\odot}$] & (\ref{eqn:mass})[$10^{9} M_{\odot}$] & [$g \; cm^{-3}$] &[$kpc$] & [$km \; s^{-1}$] \\
\hline
89.7  & 14.0 & $10^{-17}$ &9.95 & 200.78\footnotemark[1]  \\
89.7  & 23.3 & $10^{-19}$ &9.95 & 200.78\footnotemark[1] \\
370.2 & 117  & $10^{-21}$ &18.4 & 300\footnotemark[2] \\
\hline
\end{tabular}
\footnotetext[1]{fig.4 of \cite{Sofue2001}} 
\footnotetext[2]{see ref \cite{webpage}. Note, there is some evidence for a Keplerian declining rotation curve beyond 17 kpc \cite{Honma1997}. The mass distribution $> 22 \; kpc$ is still controversial \cite{Sofue2001}.}
\end{table}

\section{\label{sec:Conclusion}Conclusion}
Carmeli successfully predicted the accelerating universe \cite{Carmeli1996} two years before the announcements \cite{Garnavich1997,Perlmutter1997}.  His new metric has validity on the scale of the universe without assuming any dark matter \cite{Hartnett2004}. Here it is also shown that it may be the solution to the rotation curve anomaly in the outer regions of spiral galaxies.

Equations of motion have been derived from Carmeli's metric, which produce a Tully-Fisher type relation and describe the rotation curves in spiral galaxies without the need for non-baryonic halo dark matter. New equations for circular motion are discovered. A theoretical comparison is made with Milgrom's MOND phenomenology and some agreement found in the low acceleration limit. 

Based on the 5D \textit{space-time-velocity} of the Carmeli metric the assumed dark matter in galaxies to explain the anomalous rotation curves is no longer needed. This situation is similar to the case when Einstein introduced his general theory the advance of the perihelion of Mercury was explained adequately without the need for dark matter as had earlier been believed.

\end{document}